# Triple-Antidot Molecules

Naomi Mizuno, Dmitri V. Averin[*] and Xu Du[*]

Department of Physics and Astronomy, Stony Brook University, Stony Brook, NY 11794-3800, USA

**Abstract**

We report the realization and modeling of a triple-antidot "molecule" hosting three interacting quantum Hall quasiparticles, with tunnel coupling between antidots tunable via the magnetic field. The measured tunneling conductance spectrum reveals the molecular energy levels arising from the inter-antidot coupling and Coulomb interaction. A tunneling model is established which shows good qualitative agreement with experimental observations. This work lays the foundation for the realization of complex systems of antidots for quantum Hall quasiparticles with non-trivial quantum statistics.

*Corresponding authors: xu.du@stonybrook.edu, dmitri.averin@stonybrook.edu

Localization and manipulation of individual quasiparticles make it possible to build artificial quantum systems for quantum information science. Examples of such quantum systems include quantum dots [1-4], nitrogen-vacancy (NV) centers [5, 6], superconducting Cooper pair boxes [7-10], and superconducting quantum interference devices [11], among others. Quantum Hall (QH) antidots stand out as the systems that allow localizing QH quasiparticles with unique properties, such as fractional charges, which are not accessible in other structures. This makes QH antidots particularly interesting for studying anyon statistics and for topological schemes of quantum computation [12, 13]. In a QH antidot [14], the chiral edge mode encircling a small antidot has its energy quantized into discrete levels, mimicking a large, tunable artificial atom that hosts QH quasiparticles carrying the properties of the bulk QH fluid. The properties of these localized quasiparticles are accessible through charge reservoir-antidot tunneling. Experimentally, pioneering studies of QH antidots have been carried out using GaAs two-dimensional electron gas (2DEG), where the localization of integer [15, 16] and fractionally charged [17] quasiparticles has been demonstrated. The potential of employing such QH antidots for quantum information applications has also been discussed for some direct schemes of antidot-based quantum computing [18] or as a part of more complex quantum structures [19]. More recently, QH antidots have been demonstrated in monolayer graphene where a significantly more robust quantization effect was observed as a result of its large edge mode velocity and QH level spacing [20-22]. The ultrathin nature of graphene makes it convenient to define complex systems beyond a single QH antidot.

The ability of QH antidots to localize and arrange individual anyons into one-dimensional (1D) arrays has stimulated theoretical interest in 1D anyons [23-29], and in other effects of the anyon physics [30-32]. These phenomena, and their potential applications to quantum information devices require experimental realization and detailed modeling of the structures with controlled and coherent inter-antidot tunneling. While prior works (see, e.g., [16, 20, 21, 33-38]) have studied individual antidots rather extensively, progress on multi-antidot structures has been limited. There are several reports [39-41] implementing coupled double-antidot ``molecules'' in the GaAs-based 2D electron gas in certain specific regimes either directly [39], for the disorder-formed second antidot [41], or through an additional smaller antidot between the two main antidots [40]. However, the goal of achieving general controlled tunnel coupling of antidots as required for the localization and manipulation of individual quasiparticles remains challenging.

In this work, we demonstrate a magnetically tunable triple-antidot molecule (TAM) consisting of three tunnel-coupled QH antidots. From weak to strong magnetic field, the inter-antidot coupling decreases, resulting in the system smoothly evolving from a large single-antidot "atom" to a TAM. A complex spectrum of tunneling conductance peaks which evolve with gate-induced charge doping and magnetically tunable tunneling strength is observed. We establish a quantum mechanical model of three tunnel-coupled energy levels which achieves qualitative agreement with experimental observations.

The basic scheme of the TAM is illustrated in Figure 1a. Three antidots are placed in-line with the nearest neighbor's edge-to-edge distance small enough to allow inter-antidot tunneling. The center antidot is weakly coupled to the source and drain charge reservoirs, while the side antidots couple directly only to the center antidot. The inter-antidot and the antidot-to-reservoir coupling strengths are tuned by the magnetic field which modulates the spatial extension of the confined QH edge modes through the magnetic length ($l_B = \sqrt{\frac{\hbar}{eB}}$). A stronger magnetic field leads to a smaller $l_B$, hence tighter confinement of the edge mode and weaker inter-antidot tunneling.

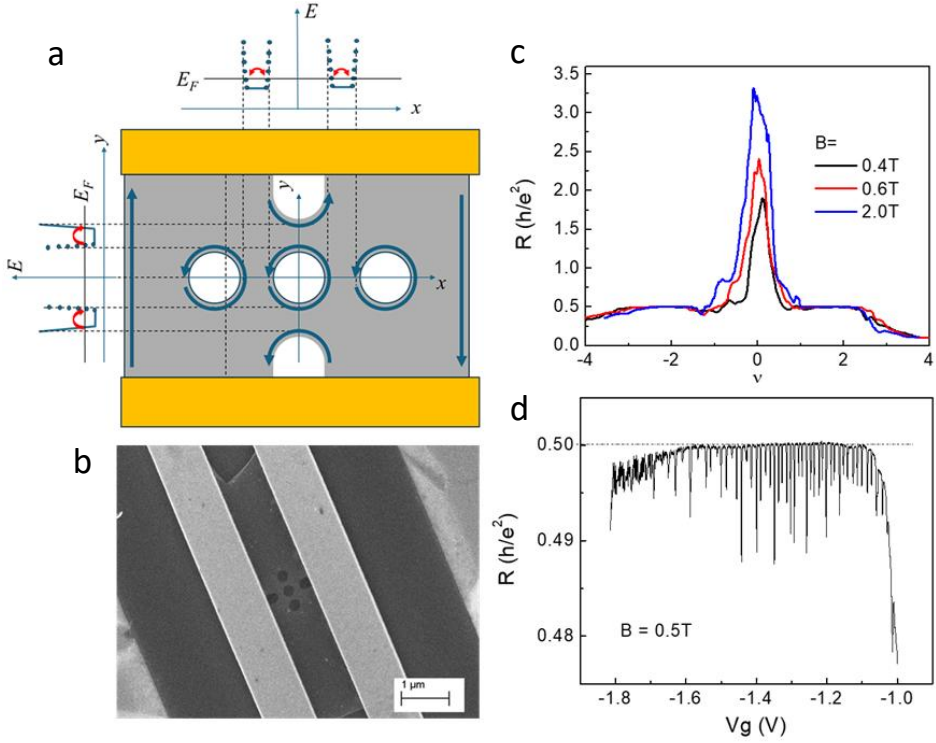

*Figure 1. Basic principles of the TAM. a) Schematic of a TAM device. Three antidots are placed in line with a small edge-edge separation. Each antidot hosts discrete quantization energy levels. Antidot-antidot and coupler-center antidot tunnelings are indicated by the red arrows. The chemical potential of the system is capacitively tuned by a back gate. b) Scanning electron microscope image of the suspended graphene TAM sample S1. c. Well-developed QH plateaus in sample S1 in indicated magnetic fields. d. Slow gate scan of the $\nu = -2$ QH plateau in 0.5 T magnetic field shows a series of resistance valleys that are associated with the TAM tunneling in sample S1.*

The samples studied in this work are patterned suspended graphene field effect devices (see SI). Figure 1b shows the scanning electron microscope image of sample S1, where the center antidot has a diameter $D_{dot} \sim 220$ nm and the outer two antidots have a slightly larger diameter of $D'_{dot} \sim 230$ nm. Here $D_{dot}, D'_{dot} \gg l_B$ so that QH edge modes are effectively confined by the antidots. In sample S1, $D_{dot}$ and $D'_{dot}$ are made slightly different so that the single-antidot energy levels of the center and the side antidots are slightly different, allowing the relative position of the energy levels to be adjustable through shifting the chemical potential. We also study the case where all three antidots have the same diameter of ~200 nm in a different sample S2, where the relative position of the energy levels is fixed. More detailed information on S2 can be found in the Supplementary Information (SI). To facilitate source/drain coupling through the center antidot, two oval-shaped notches ("couplers") are added to the system which guide the edge modes from the source/drain electrodes to ~130 nm vicinity of the center antidot edge. The center antidot-coupler's edge-edge distance is significantly larger than the nearest-neighbor antidot edge-edge distance (~90nm), allowing the TAM to be treated as a nearly isolated system, which simplifies the modeling as discussed later.

The high quality of the sample is manifested by the fully developed QH plateaus in very low magnetic field $B$, exemplified by Figure 1c. From the gate voltages ($V_g$) at the centers of the resistance plateaus ($R = \frac{h}{fe^2}$) where filling factors are $v = \frac{nh}{eB} = \frac{CV_g h}{e^2 B} = f$ ($f$ is the number of edge modes, $e$ is the electron charge, $h$ is the Planck's constant, and $n$ is the carrier density), the area gate capacitance can be estimated to be $C \approx 3.5 \times 10^{-5}$ F/m$^2$. Focusing on the $f = -2$ QH plateau and ramping the gate voltage at a very slow speed, sharp resistance dips become evident (Figure 1d), corresponding to adding single charges onto the antidots [20, 21]. From the deviation of the resistance from the $f = -2$ plateau resistance of $\frac{h}{2e^2}$, the tunneling conductance through the TAM can be calculated as $G(V_g) = \frac{1}{R(V_g)} - \frac{2e^2}{h}$.

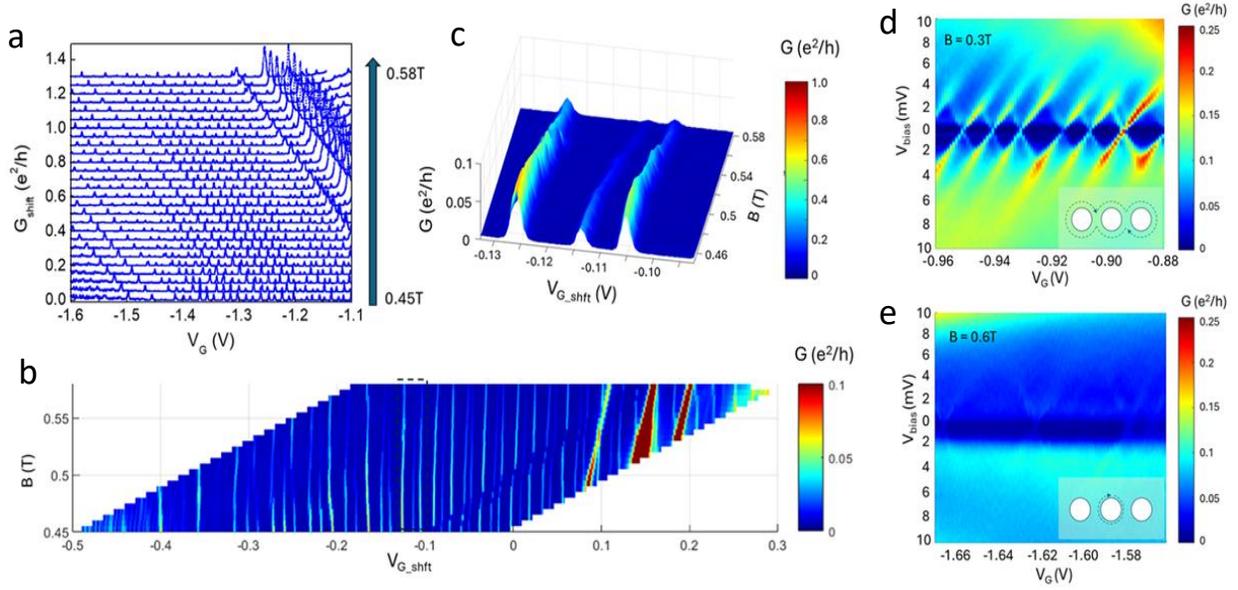

*Figure 2. Magnetic field evolution of the TAM tunneling conductance peaks in sample S1. a) Gate dependence of the TAM tunneling conductance in various magnetic fields. For clarity, the curves are shifted in the y-axis. b) Color-coded tunneling conductance versus magnetic field and shifted gate voltage $V_{G\_shft} = V_G + 2.44B$. c) Magnetic field evolution of three example conductance peaks highlighted by the dotted line box in b). d) In the strong coupling limit, the edge modes of individual antidots intersect and merge, effectively resulting in one large antidot. The corresponding charge stability diagram shows Coulomb diamonds with a small gate period and a small charging energy. e) In the weak coupling limit, the inter-antidot coupling diminishes. The charge stability diagram shows faint Coulomb diamonds with a large gate period and a large charging energy, similar to that form a single center antidot.*

In a QH antidot, each QH edge mode mimics "particle in a ring" with single-electron quantization energy levels described by $\varepsilon_j = \frac{2j\hbar v}{D} + \frac{ev\Phi}{\pi D}$, where $D$ is the effective diameter of the orbit, $v$ is the group velocity of the edge mode, $\Phi$ is the magnetic flux through the orbit, and $j$ is an integer [20, 21]. Changing the magnetic field results in the entire manifold of the energy levels shifting linearly by $\Delta j = \frac{\Phi}{\Phi_0}$, i.e., one energy level per mode per $\Phi_0$, where $\Phi_0 = \frac{h}{e}$ is the magnetic flux quantum. This scenario holds true in the presence of Coulomb interaction, but with a Coulomb gap opening at the Fermi level while the underlying energy levels shift on $\Phi$. The Coulomb interaction effectively breaks the degeneracy between QH edge modes, allowing only single-charge occupancy for each energy level, resulting in a shift of $f$ tunneling conductance peaks per $\Phi_0$ on a QH plateau with $f$ modes. Since each conductance peak corresponds to adding a single charge, the corresponding gate voltages of the conductance peaks shifts with changing magnetic field by $\frac{\Delta V_G}{\Delta B} = \frac{|f|e^2}{hC}$. The magnetic field-induced gate shift in the charge oscillation manifests the quantum coherent nature of the QH antidot modes and differentiates the QH antidot from the conventional quantum dot, despite the fact that in both systems charge transport is dominated by the Coulomb blockade [21].

Figure 2a shows the magnetic field and gate voltage dependences of the TAM tunneling conductance in sample S1. The tunneling conductance peaks show a complex gate-dependent pattern in the inter-peak spacing and peak amplitude, which shifts linearly and smoothly evolve under changing magnetic field from 0.45 T to 0.58 T. The magnetic field-induced shift in the tunneling conductance peaks is a direct manifestation of the QH antidot's energy spectrum $\varepsilon_j = \frac{2j\hbar v}{D} + \frac{ev\Phi}{\pi D}$, with a shift rate of 2 levels per magnetic flux quantum on the $f = 2$ plateau. By inverse shifting the field-dependent curves following $\frac{\Delta V_G}{\Delta B} = 2.44$V/T ratio, all conductance peaks form nearly vertical traces as shown in Figure 2b. This ratio is approximately consistent with $\frac{\Delta V_G}{\Delta B} = \frac{|f|e^2}{hC} \approx 2.2$ for our sample on the $f = -2$ QH plateau. The mild discrepancy may come from the uncertainty in the area gate capacitance.

From Figure 2b, the magnetic field evolution of the gate-dependent conductance peaks can be clearly traced. At the lower end of the applied magnetic fields the conductance peaks generally show large and similar amplitudes. The charge stability diagram shows clear Coulomb diamonds with small charging energies ≈ 2 meV (Figure 2d). With an increasing magnetic field, some of the conductance peaks show a rapid decrease in amplitude and nearly vanish at the higher end of the magnetic field, while other conductance peaks are less susceptible to the increase in magnetic field (Figure 2c). At the higher end of the applied magnetic field, the TAM tunneling conductance peaks show a complex pattern with large contrast between the peaks' heights. In the charge stability diagram, Coulomb diamonds show poor visibility because of weak reservoir-TAM tunneling. In the example shown on Figure 2e, the two large Coulomb diamonds are associated with two relatively strong conductance peaks which are separated by two weak, nearly invisible peaks. The large Coulomb diamonds indicate a large charging energy of $\approx 8 - 9 meV$. Finally, when the magnetic field reaches ~1T, there is no observable coupling to the antidot and fully flat QH plateaus are recovered.

The TAM tunneling spectrum under the lower and the higher limits of the magnetic field can be intuitively understood with a simplified scenario considering charge distribution under extreme coupling conditions. In a very low magnetic field where QH edge modes on the antidots are widely extended, overcoupling between the antidots modes effectively merge them into one large antidot. Consequently, tunneling conductance is expected to be approximately periodic in doping. And the large merged antidot size leads to small Coulomb energy, as measured by the small Coulomb diamonds. In a strong magnetic field, on the other hand, the QH edge modes encircling each antidot become tightly confined to their corresponding edges, causing the antidots to nearly decouple from each other. The tunneling conductance is therefore mainly affected by the center antidot which has about 1/3 of the size of the TAM. Indeed, the measured Coulomb energy associated with the large Coulomb diamond in Figure 2e is consistent with the theoretical estimation of the substrate-screened on-site Coulomb energy of the center antidot $\bar{u} \simeq$ 8.4 meV (see SI).

Beyond the extreme coupling limits, we next focus on the evolution of tunneling conductance through the TAM in the mid-coupling regime. In suspended graphene with minimal screening, the large on-site Coulomb energy gap dominates over other energy scales and prohibits double occupancy of charges on the same antidot. Provided that the inter-antidot coupling energy is much smaller than the Coulomb gap, only empty and singly occupied energy levels per antidot need to be considered. The basis states associated with four total TAM charge occupation values ($Q$) can be labeled by the charge numbers on the three antidots. An empty system ($Q = 0$) corresponds to the state $|000\rangle$; with a single charge, there are three basis states: $|100\rangle$, $|010\rangle$ and $|001\rangle$; with two charges there are also three basis states: $|101\rangle$, $|110\rangle$ and $|011\rangle$; and with three charges there is a single state $|111\rangle$. We set up the Hamiltonians for the states with different total charges occupation numbers:

$$Q = 0: H_0 = 0$$

$$Q = 1: H_1 = \begin{pmatrix} \varepsilon^{(1)} & t & 0 \\ t & \varepsilon^{(2)} & t \\ 0 & t & \varepsilon^{(3)} \end{pmatrix}$$

$$Q = 2: H_2 = \begin{pmatrix} \varepsilon^{(1)} + \varepsilon^{(3)} + u_{13} & t & t \\ t & \varepsilon^{(1)} + \varepsilon^{(2)} + u_{12} & 0 \\ t & 0 & \varepsilon^{(2)} + \varepsilon^{(3)} + u_{23} \end{pmatrix}$$

$$Q = 3: H_0 = \varepsilon^{(1)} + \varepsilon^{(2)} + \varepsilon^{(3)} + u_{12} + u_{23} + u_{13}$$

Here $\varepsilon^{(i)}$ ($i = 1,2,3$) is the single-particle energy of the left, center and right antidots. $u_{ij}$ is the Coulomb energy between the charges on the i-th and the j-th antidots. For $Q = 1$ and $Q = 2$, the presence of inter-antidot coupling $t$ leads to the eigenstates which are superpositions of the basis states. The general forms of the eigenstates for $Q = 1$ and $Q = 2$ are $\psi_1 = c_1^{(1)}|100\rangle + c_1^{(2)}|010\rangle + c_1^{(3)}|001\rangle$ and $\psi_2 = c_2^{(1)}|101\rangle + c_2^{(2)}|110\rangle + c_2^{(3)}|011\rangle$, respectively.

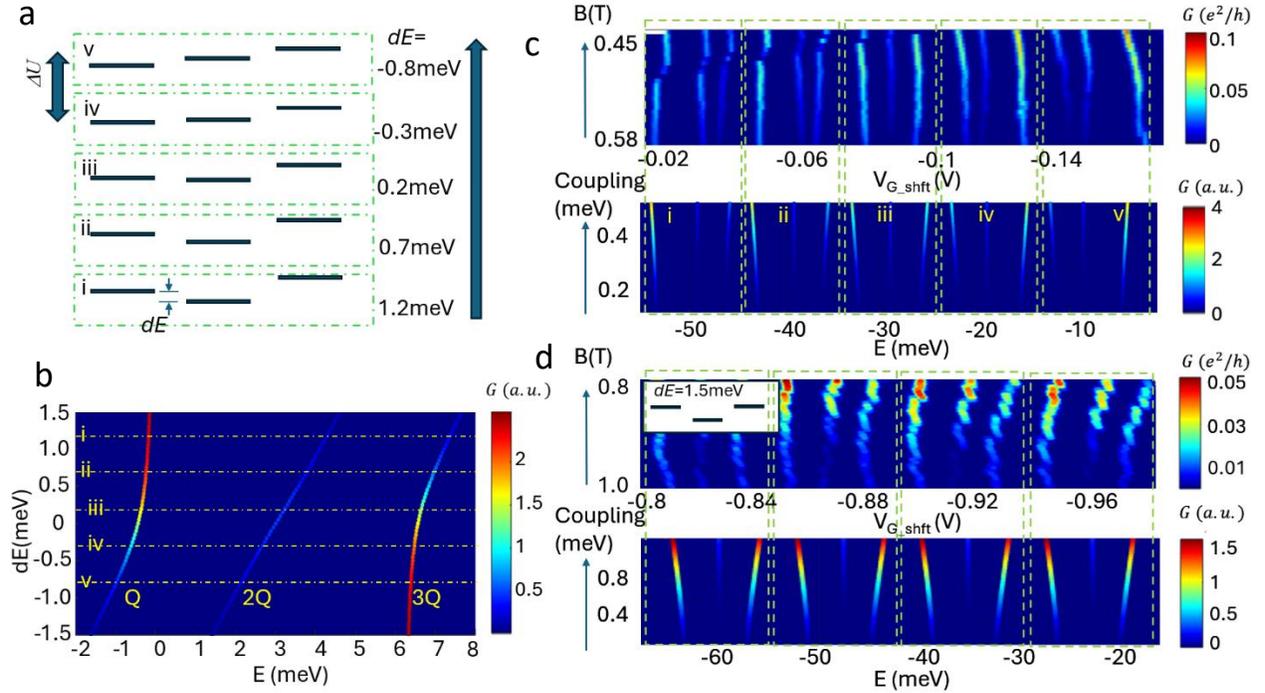

Figure 3. Modeling the TAM in the intermediate coupling regime. a) Single-particle energy levels of the three antidots in sample S1. Here the outer antidots have the same energy level spacing which is slightly smaller than that of the center antidot; and the energy levels of the right side antidot are shifted up by 2meV compared to that of the left side antidot. With changing chemical potential, the energy level configuration of the three antidot evolves, characterized by the left-center energy offset dE. b) Energies of the three tunneling conductance peaks which correspond to tunneling of the 1st, 2nd and 3rd charges on the TAM, and their dependence on dE. c) Comparison between the modeled and measured conductance peaks versus energy and

*coupling strength for sample S1. In the measurement the coupling strength is tuned by the magnetic field. The arrows on the left side of the plots point to the direction with increasing coupling strength. Note that in both plots the particle number increases from left to right. d) Comparison between the modeled and measured conductance peaks versus energy and coupling strength for sample S2. The arrows on the left side of the plots point to the direction with increasing coupling strength.*

With $\varepsilon^{(i)}$ simultaneously tuned by the gate voltage, the zero-bias voltage conductance peaks emerge when: 1) two ground states, with their corresponding $Q$ values differing by one, become degenerate; 2) the compositions of these two ground states contain basis states with the center antidot occupation number differing by one. The $Q = 0 \rightarrow 1$ tunneling conductance is associated with the transition of $|000\rangle \leftrightarrow |010\rangle$ and therefore is proportional to $\left|c_1^{(2)}\right|^2$. The $Q = 1 \rightarrow 2$ tunneling conductance is associated with the transitions including $|100\rangle \leftrightarrow |110\rangle$ and $|001\rangle \leftrightarrow |011\rangle$ and thus is proportional to $\left|c_1^{(1)}c_2^{(2)} + c_1^{(3)}c_2^{(3)}\right|^2$. And the $Q = 2 \rightarrow 3$ tunneling conductance is associated with the transition of $|101\rangle \leftrightarrow |111\rangle$ and is proportional to $\left|c_2^{(1)}\right|^2$.

With the model, energies corresponding to tunneling the 1st, 2nd and 3rd charge onto the TAM and the corresponding tunneling amplitude can be calculated for each group of $\varepsilon^{(i)}$. Increasing the gate voltage to overcome the Coulomb gap of the TAM with each antidot singly occupied, the next group of $\varepsilon^{(i)}$ is reached and leads to the next group of the three conductance peaks. The modeled energy- and coupling dependence of the tunneling conductance peaks is then qualitatively compared with their measured gate- and magnetic field dependence.

In sample S1, the energy period of the three-level (hence triple-peak) groups is estimated to be $\Delta U = u + \frac{2u_{12}}{3} + \frac{u_{13}}{3} \simeq 10.7 \text{meV}$, where $u \simeq 8.4$ meV, $u_{12} \simeq 3.1$ meV and $u_{13} \simeq 0.55$ meV are the on-site, nearest neighbor and far neighbor Coulomb energy, respectively (see SI). Based on the ~5% diameter difference between the center- and the side antidot, we estimate that over a shift of $\Delta U = 10.7$ meV, the center-side antidot energy offset $dE$ will change by approximately 0.5 meV, as illustrated by Figure 3a. We also assume that the energy levels of the right side antidot are 2.4 meV higher than that of the left side antidot. As discussed below, such a shift leads to good qualitative agreement between the model and the measurement. This small energy shift between the two outer antidots may be due to potential fluctuations which is typical in graphene from unintentional residual doping.

Figure 3b shows the impact of $dE$ on the configuration of the triple conductance peaks in sample S1, with inter-antidot coupling strength fixed at 0.4meV. With $dE$ evolving from negative to positive, the middle peak (tunneling between single and double occupancy on the TAM) moves from being closer to the first peak (between zero and single occupancy on the TAM) to being closer to the third peak (between double and triple occupancy on the TAM). At the same time, the tunneling probability associated with the first (third) conductance peak decreases (increases), while tunneling probability associated with the second conductance peak is maximized when the energies of the first and the third conductance peaks are closest to each other.

Figure 3c shows the calculated conductance peaks and their evolutions over coupling strength, over a wide chemical potential span over 5 groups of triple-peaks, with corresponding energy level offsets ($dE$) highlighted by the dotted lines in Figure 3b. Both within each $\varepsilon^{(i)}$ group and across different $\varepsilon^{(i)}$ groups, our model shows good qualitative agreement with the experimental data from sample S1 over a wide doping range covering 15 conductance peaks. At the lower hole-doping end of the plot, the first triple-peak group starts with a relatively strong peak, followed by the two weaker peaks closely located with a small gate voltage/energy separation. The separation between the second and the third peak increases with increasing coupling strength. Moving on to the higher hole energy levels with increasing doping, these conductance peak patterns evolve: the first peak of each group of the three conductance peaks becomes increasingly weaker, while the third peak of each group becomes increasingly stronger. Simultaneously, the second conductance peak moves from being closer to the third peak to being closer to the first peak. Overall, the conductance peaks with the large amplitude show a slow decay with decreasing coupling strength, while the weaker conductance peaks rapidly decay with decreasing coupling strength. These complex behaviors observed in the experiment are well captured by the theoretical model.

The TAM model is further confirmed by the measurements on sample S2 where all three antidots have the same size and hence the same energy spacing. Consequently, the triple-peaks group repeats itself with little evolution over gate voltage (Figure 3d). Within each triple-peak group, the center peak rapidly diminishes with increasing magnetic field (hence decreasing coupling strength). From our model and based on the estimated Coulomb energies from the geometry of the device ($u \simeq 10.8$ meV, $u_{12} \simeq 3.1$ meV, $u_{13} \simeq 0.63$ meV and $\Delta U = u + \frac{2u_{12}}{3} + \frac{u_{13}}{3} \simeq 13$ meV), this corresponds to a level configuration where the two outer antidots have the same energy which is ~1.5 meV higher than that of the center antidot. We note that in sample S2, the gate and magnetic field dependence of the tunneling peak traces show many jumps. The jumps can be explained by the presence of an isolated extrinsic antidot-like edge mode loop in the vicinity of the TAM (see SI), and do not alter the underlying intrinsic gate- and magnetic dependence of TAM tunneling.

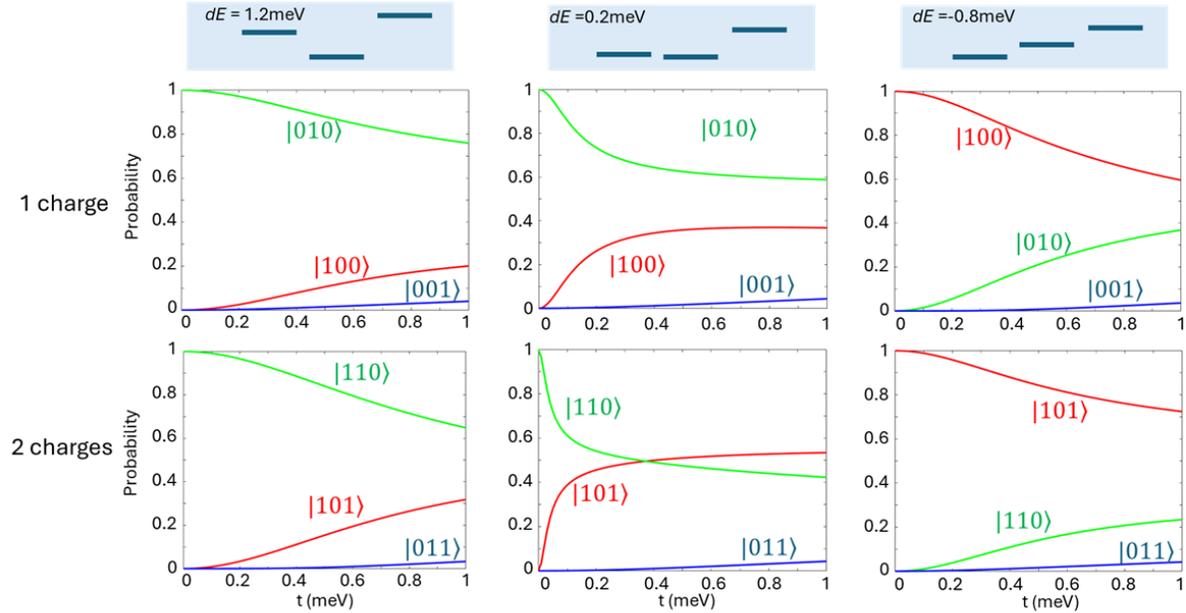

*Figure 4. Charge configurations on the TAM. From left to right, the panels correspond to the center-left antidot energy offset of $dE = 1, 0.2, -0.8$ meV, respectively. The top and the bottom panels correspond to the chare configurations of singly- and doubly-charged cases. The curves ware labeled by their corresponding basis states.*

The theoretical model provides information on the charge distribution in the TAM. As shown in Figure 4 (calculated for sample S1), under given Coulomb energies, the charge distribution depends on both the level misalignment ($dE$) and the inter-antidot coupling strength ($t$). In the presence of a single charge on the molecule and under a finite level offset between the center and the side antidots, the charge distribution has more weight on the lower energy sites, and the inter-antidot tunneling favors equilibrating the charge distribution on the three sites. With two charges on the molecule, the charge distribution is affected by both the nearest-neighbor

Coulomb repulsion and single-particle energy levels. Here again the inter-antidot tunneling tends to even out the charge distribution.

The charge distribution on the TAM helps provide a qualitative and intuitive understanding of the tunneling conductance. Within each three-peak group, the first conductance peak (associated with the $Q = 0 \to Q = 1$ transition) is determined by the weight of the charge distribution on the $|010\rangle$ state, which is the strongest when the side antidots have larger energy than that of the center antidot. As a result, the first conductance peak is observed to be strong when $dE > 0$. The second tunneling conductance peak in the three-peak group corresponds to the $Q = 1 \to Q = 2$ transition. The tunneling probability is determined by the probability of an empty center antidot ($|100\rangle$ and $|0,0,1\rangle$) in the $Q = 1$ state and an occupied center antidot ($|110\rangle$ and $|011\rangle$) in the $Q = 2$ state. As shown in Figure 4, the probability of each of these states sensitively depends on both the energy level offset and the coupling strength. Generally, a large transition probability requires at least one of the two contributing $Q = 1$ basis states ($|100\rangle$ and $|001\rangle$) and one of the two $Q = 2$ basis states ($|110\rangle$ and $|011\rangle$) to both have large amplitudes. For the combination of parameters in our model, this happens in the vicinity of $dE = -0.2\text{meV}$ shown in Figure 4, where $|100\rangle \to |110\rangle$ transition contributes significant tunneling conductance. Finally, the third conductance peak in the three-peak group is associated with the $Q = 2 \to Q = 3$ transition, with the amplitude governed by the probability of an empty center antidot in the $Q = 2$ state. From Figure 4 we can see that a relatively strong third conductance peak emerges when $dE$ becomes increasingly more negative. Under such a condition, the two charges in the $Q = 2$ state strongly favor occupying the side antidots, leaving a very low probability in the center antidot occupation which can be filled through the $Q = 2 \to Q = 3$ transition.

In conclusion, in this work we experimentally demonstrate the system of inter-coupled QH TAM in suspended graphene, where the coupling amplitude can be tuned by magnetic field strength. The energy spectrum of the system is characterized by the gate-dependent resonant tunneling conductance, which evolves over changing magnetic field as the coupling amplitude is modulated. Both the energy spectrum and its coupling dependence are studied using the quantum mechanical model, which shows good qualitative agreement with our experimental observations. This work opens the door to future studies of quantum systems of QH quasiparticles. More flexible and complex QH antidot devices may be developed with different platforms, such as the graphene/hBN heterostructures, potentially allowing independent electric field control of coupling. Beyond integer QH effect, the antidot molecules may be studied in FQH regime with fractionally charged quasiparticles and anyonic exchange statistics. On the theory side, further development may aim to understand the crossover between the weak coupling limit, the intermediate coupling regime and the over-coupling limit.

**Acknowledgement**

X.D. and D.A. acknowledges support from NSF awards under Grant No. DMR-2104781.